\documentclass[10pt]{article}
\usepackage{amssymb}
\usepackage{amsfonts}
\usepackage{bbold}
\usepackage{geometry}
\usepackage[font={small,it}]{caption}
\usepackage{amsmath}
\usepackage{amsthm}
\usepackage{mathrsfs} 
\usepackage{graphicx}
\usepackage{mathtools}
\usepackage{enumitem}
\usepackage{caption}
\usepackage[nameinlink]{cleveref}
\numberwithin{equation}{section}
\usepackage{xcolor}

\usepackage{yfonts}
\usepackage{amssymb}
\usepackage{amsmath}
\newtheorem{theorem}{Theorem}[section]

\newtheorem{remark}[theorem]{Remark}

\newtheorem{example}[theorem]{Example}

\newtheorem{property}[theorem]{Property}

\def\bearray{\begin{eqnarray}}
\def\earray{\end{eqnarray}}
\def\beq{\begin{equation}}
\def\eeq{\end{equation}}

\def\b0{{\bf 0}}

\def\bC{{\mathbb C}}           

\def\bR{{\mathbb R}}

\def\bZ{{\mathbb Z}} 

\def\gA{{\mathfrak A}}       
\def\gB{{\mathfrak B}}

\begin{document} 

\par
\bigskip
\large
\noindent
{\bf Emergent phenomena in Nature: a paradox with Theory?}
\bigskip
\par
\rm
\normalsize

\noindent {\bf Christiaan J.F.  van de Ven}\\
\par

\noindent 
Julius-Maximilians-Universit\"{a}t W\"{u}rzburg, Emil-Fischer-Stra\ss e 40, 97074 W\"{u}rzburg, Germany\\
Alexander von Humboldt Fellow \\  
Email: christiaan.vandeven@mathematik.uni-wuerzburg.de\\

\par

\rm\small

\rm\normalsize

\par
\bigskip

\noindent
\small
{\bf Abstract.} The existence of various physical phenomena stems from the concept called {\em asymptotic emergence}, that is, they seem to be exclusively reserved for certain limiting theories. Important examples are spontaneous symmetry breaking (SSB) and phase transitions: these would only occur in the classical or thermodynamic limit of underlying finite quantum systems, since for finite quantum systems, due to the uniqueness of the relevant states, such phenomena are excluded by Theory. In Nature, however, finite quantum systems describing real materials clearly exhibit such effects. In this paper we discuss these apparently ``paradoxical'' phenomena and outline various ideas and mechanisms that encompass both theory and reality, from physical and mathematical points of view. 

\normalsize

\tableofcontents

\section{Introduction}\label{introduction}

Much of modern mathematical physics concerns the search for and formalization of mathematical structures concerning  the occurrence of physical phenomena in Nature. Examples include spontaneous symmetry breaking, the theory of phase transitions, Bose-Einstein condensation, and so on. This can be done at various levels of rigor. In any event, unless perhaps in very simple cases, one can not avoid making assumptions and thus will consider a certain model that approaches reality.\footnote{The precise meaning of ``approximation of reality'' is an important point in of discussion the philosophy of physics \cite{Ardourel,Norton,Wu}. Our starting point is the one due to Butterfield \cite{Butterfield}, clarified below.} This for example already happens in the study of the hyperfine structure of the spectrum of the Hydrogen atom, where one often ``neglects'' the relativistic interactions which are strictly speaking always present. Depending on the purpose of the study such assumptions are usually well founded and therefore they (often) do not affect the description of the physical system of interest.

However, describing Nature by means of a suitable theoretical model does not always have to give a correct result consistent with experimental predictions: it can create a serious mismatch between theory and reality. A particular case for which this mismatch leads to significant ambiguity is the manifestation of {\em  emergence}, that is, according to Butterfield's first claim \cite{Butterfield}, a theoretical framework that describes  phenomena as having behavior that is new and robust with respect to some comparison class.\footnote{We also refer to the recent book by Palacios \cite{Palacios} for an overview of the topics of emergence and reduction in the philosophy of physics.} In this paper we rather focus on asymptotic emergence in the context of limits, in which emergent behavior occurs in a ``higher-level'' theory being a limit of a sequence of ``lower-level'' theories, typically as some parameter, e.g. $N$ corresponding to the size of a lattice describing quantum statistical mechanics goes to infinity, or in the spirit of the classical limit, Planck's constant $\hbar$ to zero. 
This  concept allows studying emergent phenomena  inherent in nature in a mathematically rigorous way, namely  by the theory of continuous field of $C^*$-algebras and in view of the classical limit, by strict deformation quantization \cite{Lan17}.  

On the one hand, we emphasize that this notion of emergence may be viewed as a form of reduction (in the sense of deduction), in  that, considering $N \to \infty$  (resp. $\hbar\to 0$)  enables one to deduce novel and robust behavior compared to that of systems described for finite $N$ (resp. non-zero $\hbar$). This is confirmed by two important examples one encounters in physics, namely spontaneous symmetry breaking (SSB) and phase transitions. Indeed, theoretically speaking SSB  and phase transitions can only show up in classical or infinite quantum theories seen as a classical, respectively thermodynamic limit of an underlying finite quantum theory; and these features can be deduced after taking the limit of some parameter, explained in more detail in $\S$\ref{SSBinaqt} (see also \cite{LMV,MorVen2,Ven2020,Ven2022}).

 
On the other hand, as we know from decades of experience and experimental results, emergent phenomena such as SSB or phase transitions are observed in real materials, and therefore in {\em finite} systems despite the fact that due to uniqueness principles in Theory, they seem forbidden in such systems. This does therefore not show that the limit at $N=\infty$  is ``physically real''.  The solution to this paradox is based on {\em Butterfield}'s second claim  \cite{Butterfield}:\\\\ ``{\em There is a weaker, yet still vivid, novel and robust behavior that occurs before we get to the limit, i.e. for finite $N$.  And it is this weaker behavior which is physically real.}'' 
\\\\
This can be interpreted in the sense that a robust behavior of the physical phenomenon should occur {\em before} the pertinent limit, i.e. in systems described with a finite (respectively, non-zero) parameter, meaning large but finite $N$ in the case of quantum spin systems or tiny but positive values of Planck's constant $\hbar$ in the case of classical systems. This principle,  what we call Butterfield's Principle,  is taken to be {\em our} definition of emergence.

This paper is structured in the following way. In Section \ref{SSBandphtr} we introduce the concept of phase transitions and spontaneous symmetry breaking, first from a more physical point of view often used in condensed matter physics followed by a mathematical approach common in algebraic quantum theory. In Section \ref{algebraicfor} the algebraic formalism suitable to describe classical and quantum theories is introduced. Using this language in $\S$\ref{algclaslim} the definition of the classical limit is given. In $\S$\ref{SSBinaqt} the connection with asymptotic emergence is explained and two examples (Example \ref{schroedingerexample} and Example \ref{Curieweissexample}) are provided. In Section \ref{SSBinNature} the relation between SSB in Theory versus Nature is discussed. A solution to the above paradox compatible with Butterfield's Principle is given by a mathematical mechanism and is illustrated by means of numerical simulations. Finally, in $\S$\ref{Measurement problem} the analogy between our approach and the measurement problem is outlined. The paper is concluded by posing some open problems.

\subsection{Phase transitions and spontaneous symmetry breaking }\label{SSBandphtr}
Phase transitions are everywhere in nature. They lay the foundation of many physical phenomena, like the formation of Bose-Einstein condensates, superconductivity of metals and so on. The general belief behind a phase transition is the occurrence of discontinuities in thermodynamic functionals and singularities in response functions. From a theoretical point of view these only arise in the relevant limit of an underlying finite quantum theory. 



Practically speaking, incipient discontinuities of the relevant thermodynamic functionals are already observed in experiments, e.g. as the number of particles increases, most of the change in the magnetization occurs more and more steeply, i.e. it occurs in a smaller and smaller interval around the applied field being zero \cite{Lavis}. Thus the magnetic susceptibility, defined as the derivative of magnetization with respect to magnetic field, is, in the neighborhood of zero, larger for larger $N$, and tends to infinity as $N\to\infty$. However, at finite $N$ no such singularity is present. This confirms Butterfield's Principle: the singularities which appear from experiment to exist in finite systems are not in fact singularities, but correspond to ``robust behavior'' encoded by steep maxima (incipient discontinuities) that precede the singular behavior the theory demonstrates in infinite systems. 

Another more abstract picture, common in algebraic quantum theory to study phase transitions is based on the celebrated Kubo-Martin-Schwinger (KMS) condition describing thermal equilibrium at fixed inverse temperature $\beta$ \cite{HHW}. A phase transition is now reformulated by saying that the associated $\beta$-KMS states are non-unique, that is the coexistence of more than one $\beta$-KMS state for a given dynamics. At positive temperature, however, KMS states are always unique for any finite system: no phase transitions take place \cite{BR2}. This rather abstract point of view is the usual one in modern mathematical physics; and it applies in general, i.e., it is not model-dependent. There is however also a drawback: this approach studies properties of states, and generally does not consider steep maxima. Consequently, slowly emerging singular behavior is  not taken into account and therefore it is not clear at all how to restore Butterfield's Principle. We will come back to this point in Section \ref{SSBinNature}.
\\\\
Like phase transitions, symmetry and symmetry-breaking effects affect numerous areas and aspects of mathematical and theoretical physics, e.g. it plays a role in the Higgs mechanism  providing mass generation of $W$ and $Z$ bosons, i.e. the gauge bosons that mediate the weak interaction \cite{Ruelle}.
The general and common concept behind spontaneous symmetry breaking, originating in the field of condensed matter physics where one typically considers the limit of large particle numbers, often also called thermodynamic limit, is based on the idea that if a collection of quantum particles becomes larger, the symmetry of the system as a whole becomes more unstable against small perturbations \cite{Lan13,Wez2}. 
A similar statement can be made for quantum systems in their classical limit, where sensitivity against small perturbations now should be understood to hold in the relevant semi-classical regime, meaning that a certain parameter (e.g. $\hbar$) approaches zero\footnote{The physical meaning of the limit in Planck’s constant has several interpretations. Mathematically, it is an element of the base space corresponding to a continuous $C^*$-bundle \cite{Dix}. In the context of the classical limit,  the zero-limit of this parameter corresponds to a classical theory, encoded by a commutative $C^*$-algebra. The general and correct formalism to deal with the classical limit is called {\em strict deformation quantization}, also called {\em $C^*$-algebraic deformation quantization}.} \cite{Lan17,MorVen2}.

To show the occurrence of SSB in a certain particle system various approaches of different levels of rigor are used. There are however some differences between the mathematical physics approach to SSB and the methods used in theoretical and  condensed matter physics. In the latter approaches, based mainly on the ideas of Bogoliubov, the concept that the system in question becomes sensitive to small perturbations is generally implemented by  adding a so-called infinitesimal symmetry-breaking field term to the Hamiltonian, from which one then wants to show that the limit of large particle numbers or the thermodynamic limit becomes ``singular'' \cite{Batter,Berry}, at least at the level of states, e.g. the ground state. If this happens, one says that the symmetry of the limiting system is spontaneously broken. To illustrate this we consider the quantum Ising Hamiltonian $H_{N}^{QI}$ with $N$ sites, i.e.
\begin{align}
H^{QI}_{N}: &  \underbrace{\mathbb{C}^2 \otimes \cdots  \otimes\mathbb{C}^2}_{N \: times}  \to 
\underbrace{\mathbb{C}^2 \otimes \cdots  \otimes\mathbb{C}^2}_{N \: times}; \\
H^{QI}_{N} &=- J\sum_{j=1}^{N} \sigma_3(j)\sigma_3(j+1) -B \sum_{j=1}^N \sigma_1(j). \label{CWham}
\end{align}
Here $\sigma_k(j)$ stands for $I_2 \otimes \cdots \otimes \sigma_k\otimes \cdots \otimes I_2$, where $\sigma_k$ denotes the $k$-th spin Pauli matrix ($k=1,2,3$) occupying the $j$-th slot, and  $J,B \in \mathbb{R}$ are given constants defining the strength of the spin-spin coupling and the (transverse) external magnetic field, respectively. 
The Hamiltonian  $H_N^{QI}$ is invariant under $\mathbb{Z}_2$- parity symmetry. For finite $N$, it is a well-known fact that the ground state eigenvector is (up to a phase) unique and therefore $\mathbb{Z}_2$ -invariant: no SSB occurs.
In order to observe  SSB in the limit $N\to\infty$ we take $B\in (0,1)$ and the constant $J$ then may be chosen as one. The symmetry-breaking term is typically taken to be
\begin{align}
\delta_{1/N}^{QI}=\epsilon\sum_{x=1}^N\sigma_3(x).
\end{align}
In this approach one argues that the correct order of the limits should be $\lim\epsilon\to 0 \lim N\to\infty$ \cite{VGRL18,Wez2}, which gives SSB by one of the two pure ground states on the quasi-local algebra $\mathfrak{A}\cong M_2(\mathbb{C})^{\otimes \infty}$, one of them defined by tensor products of $|\uparrow \rangle$ and the other by tensor products of $|\downarrow \rangle$, where $\sigma_3|\uparrow\rangle=|\uparrow\rangle$ and $\sigma_3|\downarrow\rangle=-|\downarrow\rangle$. The
the sign of $\epsilon$ then determines the direction of symmetry breaking (i.e. which of the limiting states is produced). In contrast, the opposite order $\lim N\to\infty \lim \epsilon\to 0$ gives a  symmetric but mixed  ground state on $\mathfrak{A}$. The symmetry is then said to be broken spontaneously if there is a difference in the order of the limits, as exactly happens in this example. In this sense, if SSB occurs, the limit $N\to\infty$ can indeed be seen as singular \cite{Batter,Berry}. However, this approach gives no insight in the physical mechanism in SSB and furthermore challenged in the philosophical literature of physics \cite{Butterfield}. Without going into the mathematical details this paper instead is based on a definition of SSB that is standard in algebraic quantum theory, i.e. it is encapsulated in terms of an algebraic formulation of symmetries and ground states \cite{BR2,Lan17}. The advantage of this approach is that it equally applies to finite and infinite systems, and to classical and quantum systems, namely it states that the ground  (or equilibrium) state, suitably defined of a system with $G$-invariant dynamics (where $G$ is some group, usually a discrete group or a Lie group) is either pure but not $G$-invariant, or $G$-invariant but mixed.\footnote{Even though at first sight both approaches to SSB might seem different, it turns out that in some cases the former approach is related to the algebraic one (see e.g. \cite[Prop. 1.6, Lemma 1.7]{DU} and \cite[Ch. 6]{Ruelle0}). Whichever approach to symmetry breaking one chooses, each has its pros and cons and which method is chosen depends on the specific purpose.} 
Subsequently, in the spirit of the above discussion the ``singularity'' of the thermodynamic limit of systems with SSB is the fact that the exact {\em pure} ground state of a finite quantum system converges to a {\em mixed} state on the limit system, explained in detail in \cite{LMV,MorVen2,Ven2020,Ven2022}.  The use of the word ``singularity'' here should thus not be confused with a  discontinuous or singular limit. 
All we mean is the inability to decompose the finite quantum state (pure state), where in the limit, however, this is possible. 

This definition is therefore compatible with the general belief that SSB only occurs  in the relevant limit. However, despite being mathematically correct, this approach does not confirm Butterfield's Principle, nor does it achieve the main idea that spontaneous symmetry breaking must be related to instability and sensitivity of the system to small perturbations in the relevant regime, manifested by the presence of non-invariant pure states rather than the unstable or non-physically mixed state. It is precisely the aim of this article to provide a natural mechanism that is mathematically correct and physically relevant, combining these (seemly different) notions of symmetry breaking into a self-contained theory.

\section{Algebraic formulation}\label{algebraicfor}
To answer the paradox in a mathematically rigorous way we need to introduce a $C^*$-algebraic framework necessary to describe states and observables. To this end we focus on the interplay between classical and quantum physics. The major advantage of this algebraic approach is to have a unified rigorous formalism suitable for both classical and quantum theories. Indeed, classical and quantum physics turn out to share the following mathematical structure:
\begin{itemize}
\item The observables constitute the (self-adjoint part of) a $C^*$-algebra $\gA$.
\item The states are positive linear functionals $\omega: \gA\to \mathbb{C}$ of norm one on $\gA$.
\item Any pure state $\omega$ has no nontrivial convex decomposition, i.e., if $\omega=t\omega_1+(1-t)\omega_2$
for some $t\in(0,1)$ and certain states $\omega_1$ and $\omega_2$, then $\omega_1=\omega_2=\omega$.
\end{itemize}
The main difference between classical and quantum observables lies in the non-commutativity of the latter.
\\\\
\noindent
{\bf Classical observables}\\
Classical observables are generally considered to be elements of the $C^*$-algebra of continuous functions vanishing at infinity on a locally compact Hausdorff space $X$, denoted by $\gA^c=C_0(X)$, where $X$ is usually a symplectic or more generally a Poisson manifold playing the role of the underlying phase space.\footnote{This choice is a matter of mathematically convenience. More options are possible to define a classical $C^*$-algebra of observables. In any case, as all commutative $C^*$-algebras are $*$-isomorphic to $C_0(X)$ for some locally compact Hausdorff space $X$, there is only one canonical classical observable algebra.} The algebraic operations are defined pointwise, i.e. $(fg)(x)=f(x)g(x)$, involution is complex conjugation, and the norm is the supremum-norm denoted by $||\cdot||_\infty$.  As a result of Riesz Representation Theorem, states $\omega:\gA^c\to \bC$ correspond to probability measures  $\mu$ on $X$. Pure states on $X$ in turn correspond to Dirac measures $\delta_x, \ (x\in X)$ defined as $(\delta_xf)=f(x)$, and therefore are in bijection with points in $X$.
\\\\
{\bf Quantum observables}\\
Quantum observables are typically elements of a non-commutative $C^*$ algebra, denoted by $\gA^q$. In the case of many situations of physical interest, $\gA^q$ usually corresponds to a a subalgebra of $B(\mathcal{H})$, the bounded operators on a certain Hilbert space $\mathcal{H}$. 
For example, $\gA^q$ could be the $C^*$-algebra of compact operators $\gA^q=\gB_\infty(L^2(\bR^n))$ on the Hilbert space $L^2(\bR^n)$, the latter denoting  the square integrable functions on $\mathbb{R}^n$. 
The algebraic relations between operators may be added and multiplied in the obvious way; involution is hermitian conjugation, and the norm is the usual operator norm. Let us for simplicity consider the case where $\gA^q=\gB_\infty(L^2(\bR^n))$. Then, states $\omega: \gB_\infty(L^2(\bR^n))\to \bC$ bijectively correspond to density matrices $\rho$ through $\rho(a)=Tr(\rho a)$. Since a unit vector $\Psi$ defines a density matrix $|\Psi\rangle\langle\Psi|$, pure states $\phi$ on $\gB_\infty(L^2(\bR^n))$  bijectively correspond to unit vectors $\Psi$ (up to a phase), via 
\begin{align}\label{algebraicstate}
\phi(a)=\langle\Psi,a\Psi\rangle.
\end{align}
A particular choice of such a unit vector is $\Psi_\hbar^{(0)}$, which corresponds to the lowest eigenvalue of a $\hbar$-dependent quantum Hamiltonian $H_\hbar$ (assuming its point spectrum is non-empty). Its corresponding algebraic state defined by \eqref{algebraicstate} is called a {\em ground state}.

\subsection{Classical limit in algebraic quantum theory}\label{algclaslim}
The theory of quantum mechanics provides an accurate description of systems containing microscopic particles, but principally quantum mechanics can be applied to any physical system. As known from long-time experience, classical physics, in turn, is a theory that works well with large objects. One may therefore expect that if quantum mechanics is applied to such objects, it reproduces classical results. Roughly speaking, this is what we call the {\em classical limit}, and refers to a way of connecting quantum with classical theories\footnote{From a $C^*$-algebraic point of view central to algebraic quantum theory, the classical limit refers to a rigorous and correct way connecting non-commutative $C^*$-algebras $\gA_\hbar$ (describing quantum theories) with commutative $C^*$-algebras $\gA_0$ (encoding classical theories) by means of convergence of algebraic states with respect to so-called quantization maps.}, rather than with infinite (non commutative) quantum systems \cite{Lan08,LanReu13}. With perhaps abuse of notation the classical limit is often denoted by $\lim \hbar\to 0$.

Let us carefully introduce this concept from an algebraic point of view starting with the famous example of phase space $X=\bR^{2n}$. We consider the {\em Schr\"{o}dinger coherent state} $\Psi_\hbar^{(q,p)}$ labeled by a point $(q,p)\in \mathbb{R}^{2n}$:
\begin{align}\label{vecPSI}
\Psi_\hbar^{(q,p)}(x) := \frac{e^{-\frac{i}{2}p\cdot q/\hbar} e^{ip\cdot x/\hbar} e^{-(x-q)^2/(2\hbar)}}{(\pi \hbar)^{n/4}} \:, \quad (x \in \bR^n\:, \hbar >0).
\end{align}
It is not difficult to prove that $\Psi_\hbar^{(q,p)}$ is a unit vector in $L^2(\bR^n, dx)$. The Schr\"{o}dinger coherent states induce a family of {\em quantization maps}, $f\mapsto Q_\hbar^B(f)$ with $f\in C_0(\bR^{2n})$ and $Q_\hbar^B(f)\in \gB_\infty(L^2(\bR^n))$ \cite{Ber}, defined by
\begin{align}
Q_\hbar^B(f):= \int_{\bR^{2n}} f(q,p)  |\Psi_\hbar^{(q,p)}\rangle \langle \Psi_\hbar^{(q,p)}| \frac{dq dp}{(2\pi\hbar)^n},  \quad (\hbar >0); \label{defQB}
\end{align}
If $(\omega_\hbar)_\hbar$ is a family of quantum states defined on $\gB_\infty(L^2(\bR^n)$ and indexed by $\hbar$ (with $\hbar\in(0, 1]$), we say that the quantum states have a {\em classical limit} if for all (or a sub-class) $f\in C_0(\bR^{2n})$, $\lim_{\hbar\to 0}\omega_\hbar(Q_\hbar(f))$ exists and defines a classical state on $C_0(\bR^{2n})$, say  $\omega_0$. 
In other words, the family of quantum states $(\omega_\hbar)_\hbar$ converges to a classical state $\omega_0$, if
\begin{align}\label{classical limit}
\lim_{\hbar\to 0}\omega_\hbar(Q_\hbar^B(f))=\omega_0(f),
\end{align} 
for all (or a sub-class) $f\in C_0(\bR^{2n})$.
An important class of states one often encounters in physics are vector states induced by eigenvectors $\psi_\hbar$ of certain Hamiltonians $H_\hbar$ parameterized by a semi-classical parameter $\hbar$.
Using these states, by Riesz' Representation Theorem the statement in \eqref{classical limit} now means that for all $f\in C_0(\bR^{2n})$ one has
\begin{align}
 \mu_0(f)=\lim_{\hbar \to 0}\int_{\bR^{2n}}f(q,p) d\mu_{\psi_\hbar}(q,p),\label{classical limit 3} 
\end{align} 
where $\mu_0$ is the probability measure corresponding to the state $\omega_0$ and $\mu_{\psi_\hbar}$, with $\hbar>0$,  is a probability measure on $\bR^{2n}$ with density $B_{\psi_\hbar}(q,p):=|\langle \Psi_\hbar^{(q,p)},\psi_\hbar\rangle|^2$ (where $\Psi_\hbar^{(q,p)}$ is the Schr\"{o}dinger coherent state vector) also called the {\em Husimi density function} associated to the unit vector $\psi_\hbar$. In other words 
$\mu_{\psi_\hbar}$ is given by
\begin{align}
d\mu_{\psi_\hbar}(q,p)=|\langle \Psi_\hbar^{(q,p)},\psi_\hbar\rangle|^2\mu_\hbar(q,p)\:,\quad  ((q,p)\in \bR^{2n})\:,
\end{align}
with $\mu_\hbar=\frac{dq dp}{(2\pi\hbar)^n}$ the scaled Lebesgue measure on $\mathbb{R}^{2n}$.
\begin{remark}
{\em 
    This approach is highly suitable for solving convergence issues regarding the semi-classical behavior of vector states induced by eigenvectors of certain operators. Indeed, such vectors itself do often not have a limit in the pertinent Hilbert space. As the next basic example already shows this problem can be circumvented by the algebraic approach introduced above.
}
\hfill$\blacksquare$
\end{remark}
\begin{example}[Classical limit of ground states of the harmonic oscillator]\label{ex:harmonicoscillator}
{\em
Consider $V(x)=1/2\omega^2x^2$ in the quantum Schr\"{o}dinger Hamiltonian (with mass $m=1/2$),
\begin{align}
H_\hbar=-\hbar^2\frac{d^2}{dx^2}+V(x),
\end{align}
densely defined on $L^2(\mathbb{R})$.
It is well known that the ground state is unique and that its wave-function $\Psi_\hbar^{(0)}(x)$ is a Gaussian peaked above $x=0$. Using the language of the above formalism, it can be shown that the sequence of algebraic ground states $\omega_\hbar^{(0)}(\cdot)=\langle\Psi_\hbar^{(0)},\cdot\Psi_\hbar^{(0)}\rangle$ converges to the Dirac measure concentrated in the single point point $(0,0)\in\bR^{2n}$. However, the ensuing limit of the eigenvector itself corresponds to a Dirac distribution which is not square integrable.

}
\hfill$\blacksquare$
\end{example}

\subsection{Spontaneous symmetry breaking and phase transitions as emergent phenomena}\label{SSBinaqt}
We shall now discuss the role of spontaneous symmetry breaking (SSB) in quantum systems and their ensuing limits.
As already elucidated in $\S$\ref{SSBandphtr}, the crucial point is that for finite quantum systems the ground state of almost any physically relevant Hamiltonian is unique\footnote{Perhaps the physically most famous  exception to this idea is the ferromagnetic quantum Heisenberg model, where the ground state for any finite $L$ is degenerate.} and hence invariant under whatever symmetry group $G$ it may have \cite{Lan13,VGRL18}. Hence, the possibility of SSB, in the usual sense of having a family of asymmetric pure ground states related by the action of $G$, seems to be reserved for infinite systems. Using the analogy between the thermodynamic limit and the classical limit (Section $\S$\ref{algclaslim}), an exact similar situation occurs in quantum mechanics of finite systems (SSB typically absent), versus classical mechanics, which allows it. 
Therefore, generally speaking, spontaneous symmetry breaking can be seen as natural {\em emergent phenomenon}\footnote{We refer to \cite{And} for a detailed discussion on this topic.} meaning that it only occurs in the limit (be it an infinite quantum or a classical system) of an underlying finite quantum theory. 


Let us give two examples of SSB occurring in the classical limit. The first example concerns a certain class of Schr\"{o}dinger operators where emphasis is put to the classical limit in $\hbar\to 0$ appearing in front of the Laplacian, the second example is the quantum Curie-Weiss model where the parameter $\hbar\to 0$ is interpreted as the limit of large number of particles $N$.

\begin{example}[Schr\"{o}dinger operator]\label{schroedingerexample}
{ \em
Consider the Schr\"{o}dinger operator $H_\hbar$ suitably defined on a dense domain of $L^2(\bR^n)$ by
\begin{align}
H_\hbar=-\hbar^2\Delta+V
\end{align}
where $\Delta$ is the Laplacian and $V$ a potential acting by multiplication.  We take $V$ to be of the form $V(x)=(x^2-1)^2$, describing particularly a double well ($n=1$) or Mexican hat ($n=2$) potential. Clearly $V$ is invariant under the natural $SO(n)$-action, and this property transfers to $H_\hbar$. Furthermore, by standard results $H_\hbar$ has compact resolvent \cite{RS4}, and as a result of the (infinite version of the) Perron-Frobenius Theorem, the ground state  eigenvector $\Psi_\hbar^{(0)}$  of the corresponding Schr\"{o}dinger operator $H_\hbar$ is (up to phases) {\em unique}. The following facts are valid.
\begin{itemize}
\item[-] The algebraic (pure) ground state $\omega_\hbar^{(0)}(\cdot)=\langle \Psi_\hbar^{(0)},\cdot\Psi_\hbar^{(0)}\rangle$ is unique and $SO(n)$-invariant. Hence, no SSB occurs for any $\hbar\neq 0$.
\item[-] In a similar fashion as in Example \ref{ex:harmonicoscillator} the classical limit $\omega_0=\lim_{\hbar\to 0}\omega_\hbar^{(0)}$ exists; it yields a $SO(n)$-invariant mixed state $\omega_0^{(0)}$, given by
\begin{align}\label{claslimsch}
\omega_0^{(0)}(f)=\int_Gf(g\sigma_0)d\mu_G(g);
\end{align}
where $\mu_G$ denotes the associated Haar measure, and $\sigma_0\in\bR^{2n}$ is an arbitrary point in the level set $\Sigma_0:=\{\sigma=(q,p) \ | \ p=0, ||q||=1\}$. In fact, the quantum Hamiltonian $H_\hbar$ on $L^2(\bR^n)$ has a classical counter part, namely the function $h_0^{Sch}$ defined on phase space $\bR^{2n}$ by
\begin{align}
h_0^{Sch}(q,p)=p^2+V(q), \ \ ((q,p)\in\bR^{2n})
\end{align}
It is precisely the set $\Sigma_0$ that corresponds to the minima of $h_0^{Sch}$, i.e. $\Sigma_0=(h_0^{Sch})^{-1}(\{0\})$. 
\item[-] Since the state $\omega_0^{(0)}$ (cf. \eqref{claslimsch})  defines a  $SO(n)$-invariant, but mixed ground state one can prove that the $SO(n)$ symmetry is spontaneously broken.\footnote{We refer to \cite{MorVen2,Ven2020} for a mathematical discussion on ground states, spontaneous symmetry breaking  and the classical limit in algebraic quantum theory, where the latter concept is  strongly related to the theory of strict deformation quantization.}
\end{itemize}

\hfill$\blacksquare$
}
\end{example}
\begin{example}[Curie-Weiss model]\label{Curieweissexample}
{ \em
Consider the quantum Curie-Weiss Hamiltonian $H_N^{CW}$ with $N$ sites
\begin{align}
H^{CW}_{N}: &  \underbrace{\mathbb{C}^2 \otimes \cdots  \otimes\mathbb{C}^2}_{N \: times}  \to 
\underbrace{\mathbb{C}^2 \otimes \cdots  \otimes\mathbb{C}^2}_{N \: times}; \\
H^{CW}_{N} &=- \frac{J}{2N}\sum_{i,j=1}^{N} \sigma_3(i)\sigma_3(j) -B \sum_{j=1}^N \sigma_1(j). \label{CWham}
\end{align}
In order to observe spontaneous symmetry breaking in the relevant limit we take  $B\in (0,1)$ and then we may choose $J=1$. The following holds.
\begin{itemize}
\item[-] The (pure) ground state $\omega_N^{(0)}(\cdot)=\langle \Psi_N^{(0)},\cdot\Psi_N^{(0)}\rangle$ is unique and $\mathbb{Z}_2$-invariant. Again, no SSB occurs for any $N<\infty$.
\item[-] Similar as in Example \ref{schroedingerexample} the classical limit $\omega_0=\lim_{\hbar\to 0}\omega_\hbar^{(0)}$ exist;\footnote{The way proving this is to use macroscopic averages which define quantization maps as well \cite{LMV,Ven2020}.} it yields a $\mathbb{Z}_2$-invariant, but mixed ground state $\omega_0^{(0)}\in C(B^3)$, given by
\begin{align}
\omega_0^{(0)}(f)=\frac{1}{2}(f(x_-)+f(x_+)),
\end{align}
where $x_\pm\in B^3$ correspond to the minima of the function $h_0^{CW}(x,y,z)=-(\frac{J}{2}z^2+Bx)$, and $B^3:=\{(x,y,z)\in \bR^3 \ |\ x^2+y^2+z^2\leq 1\}$ denotes the closed unit-ball $\mathbb{R}^3$. Again, following the algebraic definitions it is not difficult to see that the $\mathbb{Z}_2$-symmetry is spontaneously broken.
\end{itemize}
\hfill$\blacksquare$
}
\end{example}
In view of the previous discussion, we conclude that SSB shows up as ``emergent phenomenon'' when passing from the quantum realm to the classical world, i.e. by considering a classical limit,  $\hbar\to 0$ in Example \ref{schroedingerexample} and $N\to \infty$ in Example \ref{Curieweissexample}. This should be understood as emergence in the sense of deduction, as the novel and robust behavior (i.e. SSB) arises only {\em after} taking the limit in the relevant parameter.

In this algebraic fashion a similar study may be performed for phase transitions. In that case, one must show that the KMS states at inverse temperature, which are unique for finite systems, coexist in infinite systems. This is much harder problem than the study of SSB, since in general no additional symmetry is imposed on the relevant model. One usually tries to show that the quantum free energy function converges, followed by characterizing its minima and analyzing its continuity properties. This goes beyond the scope of this work, but we examine a simple yet illustrative case in $\S$\ref{phase transitions paradox}.

\section{Theory versus Nature}\label{SSBinNature}
The previous examples have shown that SSB is a natural phenomenon emerging in the limit $N\to\infty$ (where $N$ typically plays the role of the number of lattice sites) or in the context of Schr\"{o}dinger operators, in Planck's constant $\hbar\to 0$ (which may be interpreted as a limit in large mass). Indeed, a pure $G$-invariant (quantum) ground state typically converges to a mixed $G$-invariant (classical) ground state. 
For phase transitions instead this is a much harder question. We will therefore mainly focus in SSB and step-by-step show how this algebraic approach in a rigorous way leads to the fulfillment Butterfield's Principle, thereby solving the paradox (cf.  $\S$\ref{solution paradox}).

First of all, even though this notion of SSB is theoretically correct, it is not the entire story, since according to the general definitions, SSB also occurs whenever pure (or more generally, extreme) ground states are not invariant under the pertinent symmetry group $G$. It is precisely the latter notion of SSB that one observes in Nature: the limiting ground state is typically measured to be asymmetric in the sense that it is pure but {\em not} $G$-invariant. This state is therefore also called the ``physical'' ground state. Starting from this physical idea that the limiting ground state must be pure but not $G$-invariant, to prove SSB one should introduce a kind of quantum ground state that converges to a pure but not $G$-invariant, hence physical  ground state, instead of to the unphysical $G$-invariant mixture predicted by the theory. 

To get an idea of what this means we  first go back to Example \ref{schroedingerexample} for Schr\"{o}dinger operators $H_\hbar$ with a $SO(n)$-invariant potential of the form $V(x)=(x^2-1)^2, \ (x\in \bR^n)$, i.e. we consider the operator 
\begin{equation}
H_{\hbar}=-\hbar^2\frac{d^2}{dx^2}+(x^2-1)^2, \label{TheHam}
\end{equation}
The symmetry group $SO(n)$ acts on $\mathbb{R}^n$ by rotation. This induces the obvious action on the classical phase space $\mathbb{R}^{2n}$, i.e., $R(p,q) = (Rp,Rq), \ ((p,q)\in \mathbb{R}^{2n})$, as well as a unitary map $U$ on the Hilbert space  $\mathcal{H}:=\mathcal{H}_\hbar=L^2(\mathbb{R}^n)$ defined by $U_R\psi(x) = \psi(R^{-1}x)$. Furthermore, as we have seen in Example \ref{schroedingerexample}, for any $\hbar>0$ and $n\in \mathbb{N}$ the ground state of this Hamiltonian is unique and hence invariant under the $SO(n)$ action; with an appropriate phase choice it can be chosen to be real and strictly positive. Yet the associated classical Hamiltonian 
\begin{equation}
h_0^{sch}(p,q)=p^2+ (q^2-1)^2, \label{TheHamc}
\end{equation}
defined on the classical phase space $\mathbb{R}^{2n}$, has a  degenerate ground state depending on the dimension $n$. For convenience we focus on two cases, $n=1$ and $n=2$.  The case $n=1$, corresponding to the symmetric double well potential with $\bZ_2$-symmetry, the minima of $V$ are obviously two-fold degenerate and given by the Dirac measures $\omega_{\pm}$ defined by
 \begin{equation}
\omega_{\pm}(f)=f(p=0,q\pm 1), \label{voorm}
\end{equation}
where the point(s) $(p=0,q=\pm 1)$ are the (absolute) minima of $h_0^{sch}$. These (pure) states are clearly not $\mathbb{Z}_2$-invariant. Indeed, the $\bZ_2$-action on $\bR^2$ induced by he non-trivial element of $\bZ_2$ is defined by the map $\zeta:x\mapsto-x$, so that  $\omega_{\pm}\mapsto\omega_{\mp}$.

The case $n=2$ is the famous Mexican hat potential with $SO(2)$ rotation symmetry.  As we have seen before the level set $\Sigma_0=(h_0^{sch})^{-1}(\{0\})$ is the $SO(2)$-orbit through the points ($||p||=0$ and $||q||=1$.) This set of ``classical'' ground states is now connected and forms a circle in phase space.  This is the idea of the (classical) Goldstone Theorem, a particle can freely move in this circle at no cost of energy, i.e., the motion in the orbit $\Sigma_0$ is free. Similarly, since each point in phase space is rotated under the $SO(2)$ action, no point in $\Sigma_0$ is $SO(2)$-invariant.
\\\\
To explain the issue about the existence of spontaneous symmetry breaking in Nature rather than Theory, then translates to the quest for a mechanism that reproduces Dirac measures in the relevant limit. To do this, we summarize two important approaches used in literature.
\\\\
{\bf Anderson's approach }\\
A commonly used mechanism to accomplish localization, originating with Anderson (1952) \cite{And52,And}, is based on forming symmetry-breaking linear combinations of low-lying states (sometimes called {\em ``Anderson’s tower of states''}  \cite{Tasaki19}) whose energy difference vanishes in the pertinent limit. In the limit (i.e. either $\hbar\to 0$  or $N\to\infty$) these low-lying eigenstates, still defining a pure state, converge to some symmetry-breaking pure ground state on the relevant limit system. In view of the quantum Ising model introduced in section $\S$\ref{SSBandphtr} this limiting state corresponds to one of the tensor product states induced by the vectors $|\uparrow\rangle$ or $|\downarrow\rangle$.

In view of Example \ref{schroedingerexample} with $SO(n)$ symmetry, the case $n=1$ precisely yields two eigenstates, namely the ground state $\Psi_\hbar^{(0)}$ and the first excited state $\Psi_\hbar^{(1)}$ with eigenenergies $E_\hbar^{(0)}$ and $E_\hbar^{(1)}$, respectively, such that the difference 
$\Delta_{\hbar}:=|E_\hbar^{(0)}-E_\hbar^{(1)}|$ vanishes as $\hbar\to 0$. In contrast to the classical limit of the pure algebraic vector state $\omega_\hbar^{(0)}$ corresponding to the exact ground state $\Psi_\hbar^{(0)}$  (the same is true for the first excited state $\omega_\hbar^{(1)}$), viz. the mixed state $\omega_0^{0}=\frac{1}{2}(\omega_0^++\omega_0^-)$, the low-lying eigenstates defined by $\Psi_\hbar^{\pm}:=\frac{\Psi_\hbar^{(0)}\pm\Psi_\hbar^{(1)}}{\sqrt{2}}$ which are now localized, but define pure states $\omega_0^{\pm}$ as well, converge to pure classical states:
\begin{align}
&\lim_{\hbar\to 0}\omega_\hbar^{(\pm)}=\omega_0^{\pm}.
\end{align}
In this simple example this is clear, because $\Psi_\hbar^{\pm}$ has a single peak above $\pm 1$ since neither $\Psi_\hbar^{+}$ nor $\Psi_\hbar^{-}$ is an energy eigenstate (whereas their limits $\omega_0^{+}$ and $\omega_0^{-}$ are energy eigenstates, in the classical sense of being fixed points of the Hamiltonian flow). The reason is that the energy difference $\Delta_{\hbar}$ vanishes exponentially as $\hbar\to 0$, so that in the classical limit $\Psi_\hbar^{+}$ and $\Psi_\hbar^{-}$  approximately do become energy eigenstates.

A similar situation occurs for the the Mexican hat potential, i.e. $n=2$. 
As opposed to the double well potential, we now have an infinite number of low-lying eigenstates $\{\Psi_\hbar^{(n)},n\in \bZ\}$, in the sense that they are all eigenfunctions of the operator $H_\hbar$ and their eigenenergies all coincide in the limit $\hbar\to 0$. One then define the localized wave function by
\begin{align}
\Psi_{\hbar,\theta}^N=\frac{1}{\sqrt{2N+1}}\sum_{n=-N}^NU_\theta\Psi_\hbar^{(n)},
\end{align}
where $\theta\in [0, 2\pi)$ parameterizes the $SO(2)$-action on $L^2(\bR^2)$. The limit
\begin{align} 
\lim_{N\to\infty }\Psi_{\hbar,\theta}^N,
\end{align}
does not exists in $L^2(\bR^2)$. Instead, the idea is to exploit the techniques explained in $\S$\ref{algclaslim}, which makes the unit vectors $\Psi_{\hbar,\theta}^N$ converge to some
probability measure $\mu_{\hbar,\theta}$ on $\bR^{2n}$ as $N\to\infty$.  In the subsequent limit $\hbar\to 0$, one should then obtain a probability measure $\mu_{0,\theta}$ concentrated on a suitable point in the orbit of classical ground states $\{(p,q)\in\mathbb{R}^4 \ | \ p=0,||q||=1 \}$ \cite{Lan17}.

To summarize, these two examples illustrating Anderson's approach show that the right physical ground state is obtained in the pertinent limit. Clearly the pertinent pure (not $SO(n)$-invariant) limiting state breaks the $SO(n)$-symmetry. 
\subsection{Solving the paradox: case of SSB}\label{solution paradox}
Even though Anderson's approach may yield the right ``physical'' ground state  in the pertinent limit, the low-lying tower of eigenstates itself is {\em not} an eigenstate of the original operator, and breaks the pertinent symmetry already for any finite $\hbar$ (resp. any $N$). This approach therefore is rather artificial: apart from the fact that this approach forces a separated limit to be taken, moreover, it  still does not account for the fact that in Nature real and hence {\em large and finite} materials evidently display spontaneous symmetry breaking whereas on microscopic scale no SSB should be present, i.e. the pure state should be $G$-invariant in such regimes. Another more sophisticated mechanism that extends Anderson's approach and is compatible with as well theory as nature is introduced below.
\\\\
{\bf Perturbative approach} \\
In order to provide a natural mechanism compatible with symmetry breaking in Nature, we recall Butterfield's  Principle ~\cite{Butterfield}: ``There is a weaker, yet still vivid, novel and robust behavior that occurs before we get to the limit, i.e. for finite $N$. And it is this weaker behavior which is physically real.''  Despite the fact that this was originally phrased in the context of phase transitions, namely that robust behavior for finite systems is not singular but corresponds to steep maxima (what Lavis et al., call incipient singularities \cite{Lavis}), this can can be interpreted for systems exhibiting SSB as well. This means that some approximate and robust form of symmetry breaking should already occur {\em before} the pertinent limit (viz. finite $N$ or non-zero $\hbar$), despite the fact that uniqueness of the ground state seems to forbid this. To accomplish this, it must be shown that for finite $N$ or $\hbar>0$ the system is not in its exact ground state, but in some other state having the property that as $N\to\infty$ or $\hbar\to 0$, it converges in a suitable sense (see the discussion in $\S$\ref{algclaslim} and \cite{Ven2022}) to a symmetry-broken ground state of the limit system, which is either an infinite quantum system or a classical system. Since the symmetry of a state is preserved under the limits in question, this implies that the actual physical state at finite $N$ or $\hbar$ must already break the symmetry.

A natural setting taking this into account is based on some form of perturbation theory \cite{Fraser,VGRL18}. Although in a different context, this approach, firstly introduced by Jona-Lasinio et al. \cite{Jona} and later called the ``Flea on the elephant'' by Simon \cite{Sim85}, is based on the fact that the exact ground state of a tiny perturbed Hamiltonian approximates the right physical (symmetry-broken) state in such a way that this perturbed ground state is $G$-invariant for relatively large values of $\hbar>0$ (or similarly, for small values of $N$), but when approaching the relevant limiting regime (i.e. $\hbar \ll 1$ or $N \gg 0$) the perturbed ground state loses its $G$-invariance,\footnote{We remind the reader that for any $\hbar>0$ or $N<\infty$ the unperturbed ground state of a generic Hamiltonian is typically $G$-invariant, so that no SSB occurs. Therefore, strictly speaking any approach to symmetry breaking in Nature ($\hbar>0$ or $N<\infty$ ) is {\em explicit} rather than spontaneous.} and therefore breaks the symmetry already {\em before} the ensuing limit. The physical intuition behind this mechanism is that these tiny perturbations should arise naturally and might correspond either to imperfections of the material or contributions to the Hamiltonian from the environment. This is therefore also compatible with the physical idea that the symmetry of the system in question should become highly sensitive to small perturbations in the relevant regime (viz. Section \ref{introduction}).

\begin{figure}[htb!]
\centering
\includegraphics[width=10cm]{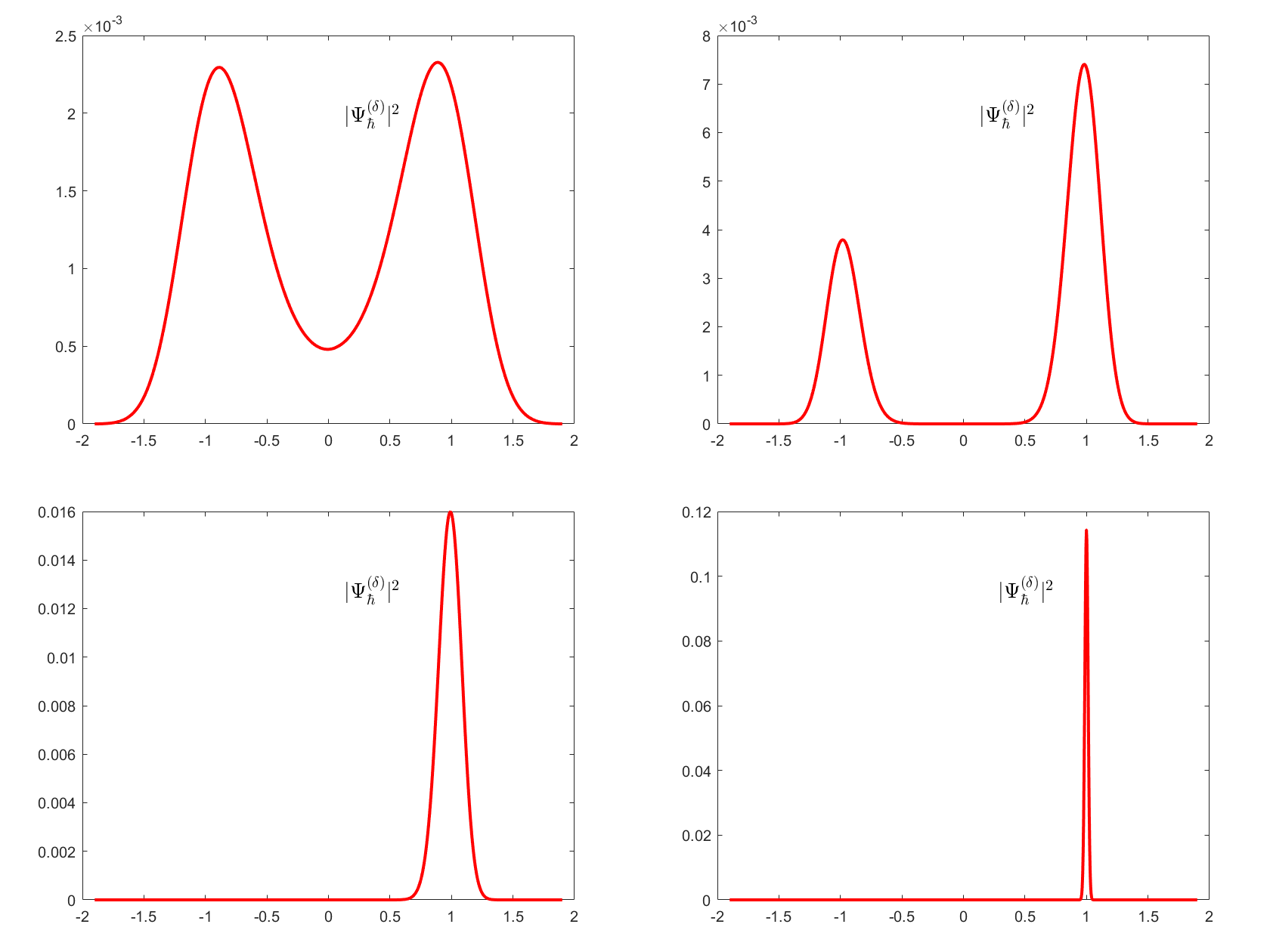}
\caption{The ground state eigenfunction $|\Psi_\hbar^{(\delta)}|^2$ of the perturbed Hamiltonian $H_\hbar^{\delta}$ corresponding to the double well potential for four values of Planck's constant $\hbar$. The figure in the left top corner corresponds to $\hbar=0.5$, the figure in the right top top to $\hbar=0.1$, the figure in the left corner down to $\hbar=0.05$, and the  figure in the right down corner to $\hbar=0.01$.}
\label{localizationdoublewell}
\end{figure}

To see what this means in a very simple context we again consider the Schr\"{o}dinger operator defined by \eqref{TheHam} for $n=1$.\footnote{To visualize the situation properly we restrict the Schr\"{o}dinger operator to $L^2([-2,2])$.} As already mentioned, for any $\hbar>0$ the (normalized) ground state $\Psi_{\hbar}^{(0)}\in L^2(\mathbb{R})$  of this Hamiltonian is unique and hence invariant under the $\mathbb{Z}_2$-symmetry $\psi(x)\mapsto\psi(-x)$; with an appropriate phase choice it is real, strictly positive, and doubly peaked above $x=\pm 1$. The classical Hamiltonian $h_0^{sch}\in C(\mathbb{R}^2)$ has a two-fold degenerate ground state, given by the Dirac measures $\omega_{\pm}$ defined by \eqref{voorm}.
As predicted by Theory, the symmetric mixed state $\omega_0=\frac{1}{2}(\omega_++\omega_-)$ turned out to be the classical limit  of the exact algebraic ground state $\omega_{\hbar}$ of \eqref{TheHam} as $\hbar\to 0$. 
In view of the previous discussions we perturb \eqref{TheHam} by adding an asymmetric term $\delta V$ (i.e., the ``flea''), which, however small it is, under reasonable assumptions localizes the ground state  $\psi^{(\delta)}_{\hbar}$ of the perturbed Hamiltonian $H_\hbar^{(\delta)}=H_\hbar+\delta V$ in such a way that $\omega^{(\delta)}_{\hbar}\to \omega_+$ or $\omega_-$ (rather than the unphysical mixture $\omega_0$) depending on the sign and location of $\delta V$.\footnote{In this approach there is therefore no need for ``low-lying eigenstates'', i.e. the exact ground state of the perturbed model already reflects the physical situation as is clearly visible from Figure \ref{localizationdoublewell}. Besides, no separate limit has to be taken since localization follows from $\hbar\to 0$.} According to the results proved in \cite{Sim85} to get localization the perturbed Hamiltonian should have the following properties:

\begin{property}\label{propertiesflealike}
\leavevmode
{\em 
\begin{itemize}
\item $\delta V$ is smooth, bounded and vanishes in a neighbourhood of the minima of $V$.
\item The support of $\delta V$ is localized at positive distance from the minima of $V$, but not too ``far'', meaning that its support is located not farther way than the Agmon distance  between the two minimal points (see \cite{VGRL18,Sim85} for technical details). 
\end{itemize}
}
\hfill$\blacksquare$
\end{property}

\begin{figure}[htb!]
\centering
\includegraphics[width=10cm,height=5cm]{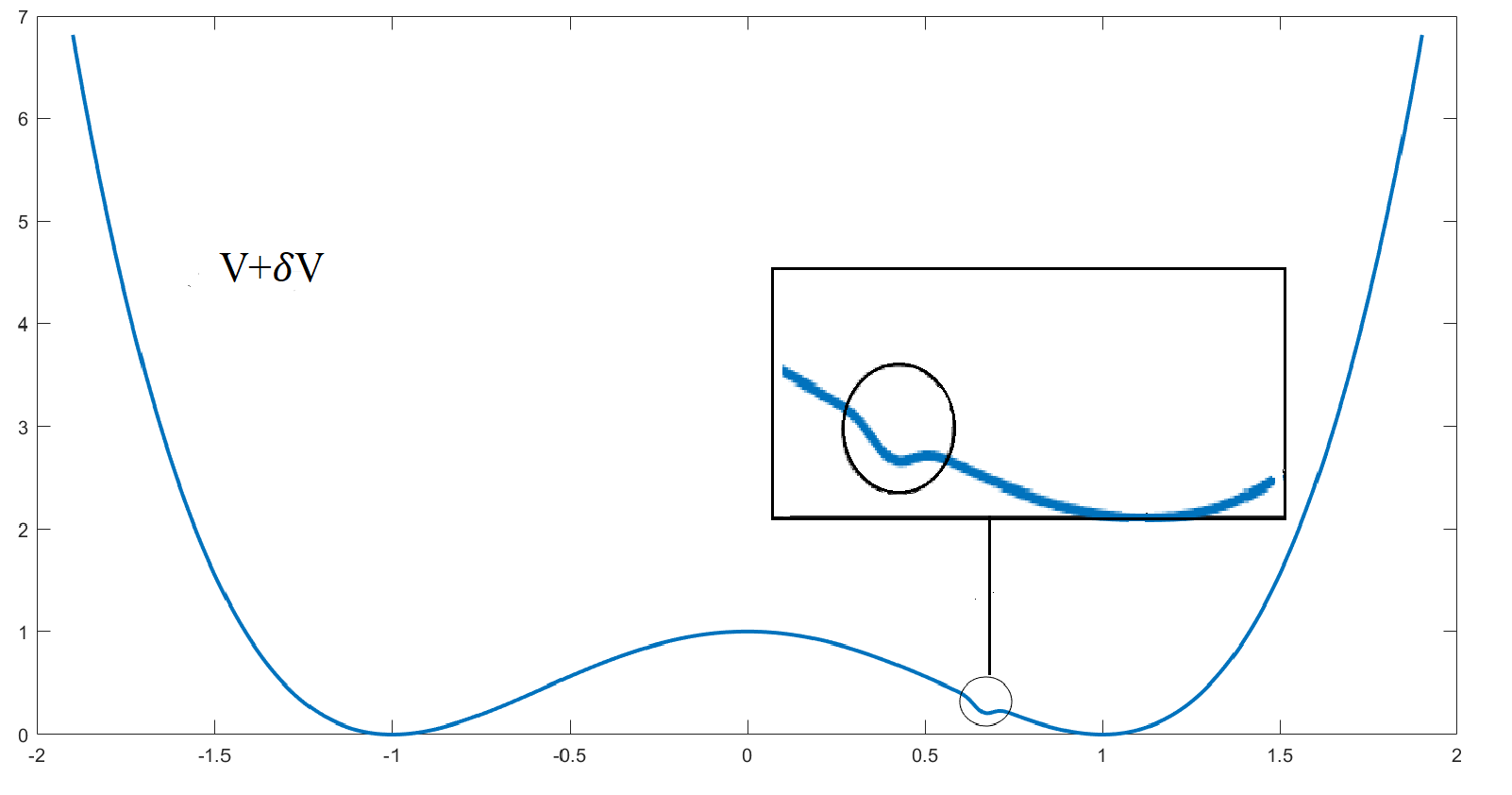}
\caption{The $\bZ_2$-symmetric double well potential $V$ with asymmetric perturbation $\delta V$ whose support is indicated by the black circle. The perturbation $\delta V_{b,c,d}$ is defined for the parameters, $b=0.65$, $c=0.2$, $d=-0.1$.}
\label{wellreuvers1}
\end{figure}

\begin{remark}\label{conditionsfleaexplained}
{\em 
The first condition of Property \ref{propertiesflealike} means that the perturbation should not have any support at the minimum set itself. The physical interpretation is that tiny perturbations should induce symmetry breaking for small values of $\hbar$, but the symmetry of the minimum (or more generally, the level) set itself should be untouched to not affect the physical situation. The second condition means that the perturbation is indeed ``effective''. For example, if the perturbation is such that the support is located at a distance larger than the distance between both minima, no effect is observed: the ground state eigenvector will remain $\mathbb{Z}_2-$invariant for any $\hbar$.
}
\hfill$\blacksquare$
\end{remark}
Let us consider an example. A natural perturbation $\delta V$ for the Schr\"{o}dinger operator $H_\hbar$ is
\begin{align}
\delta V_{b,c,d}(x)=
\begin{cases} d \exp{\bigg{[}\frac{1}{c^2}-\frac{1}{c^2-(x-b)^2}\bigg{]}},  \ \ \text{if} \  |x-b|< c; \\
0, \ \ \text{if} \ |x-b|\geq c,\label{perturbationflea1}
\end{cases}
\end{align}
where the parameters $(b,c, d)$ represent the location of its center $b$, its width $2c$ and its height $d$, respectively. Using this function as  perturbation for a suitable choice of these parameters it can be shown that indeed all conditions of Property \ref{propertiesflealike} can be met. This is confirmed in Figures \ref{wellreuvers1}-\ref{localizationdoublewell}, where  the perturbed potential  $V+\delta V$ and the ground state eigenvector of $H_\hbar^{(\delta)}$ for several values of $\hbar$ are plotted.\footnote{The ground state wave function of the perturbed Hamiltonian (which has two peaks for $\delta V=0$ ) localizes in a direction in which the perturbation assumes a relative minimum. For example, if $\delta V$ is positive and is localized to the right, then the relative energy in the left-hand part of the double well is lowered, so that localization will be to the left. If $\delta V$ is negative and localized on the right, then the  localization will be to the right, as is the case in Figure \ref{localizationdoublewell}.} From Figure \ref{localizationdoublewell} it is clear that the ground state eigenfunction already localizes for relatively small, but positive values of $\hbar$, whilst for relatively large values of $\hbar$ the state is $\bZ_2$-invariant. For such small values of $\hbar$ the ground state is definitely not  $\bZ_2$-invariant anymore, and therefore it corresponds to the physical (symmetry broken) ground state yielding a Dirac measure in the subsequent limit $\hbar\to 0$. This precisely confirms the essence of the argument and the flea mechanism: SSB is already foreshadowed in quantum mechanics for small yet positive $\hbar$.

\begin{remark}
{\em 
It has been numerically checked that for the values of $\hbar$ used to compute the ground state eigenfunction $\Psi_\hbar^{(\delta)}$ of $H_\hbar^{(\delta)}$, the ground state energy $E_0^{(\delta)}$ is indeed non-degenerate: no degeneracy effects due to numerical noise has played a role. Therefore, for these values of $\hbar$ (obviously, they depend on the machine precision of the computer used to perform calculations) the simulations accurately reflect the uniqueness of the ground state, proved from general theory on Schr\"{o}dinger operators. 
}
\hfill$\blacksquare$
\end{remark}

\newpage
\begin{remark}
{\em
We stress that this ``flea mechanism''  has been independently verified for mean-field quantum spin systems in the thermodynamic limit $N\to\infty$ in \cite{VGRL18}. In this work a suitable perturbation in spin space has been defined having the same properties as perturbation \eqref{perturbationflea1}. As a result the exact pure ground state of the perturbed quantum spin Hamiltonian converges to a pure state on the (in this case) phase space $B^3:=\{(x,y,z)\in \bR^3 \ |\ x^2+y^2+z^2\leq 1\}$.
}
\hfill$\blacksquare$
\end{remark}

\begin{figure}[!htb]
\centering
\includegraphics[width=9cm,height=7cm]{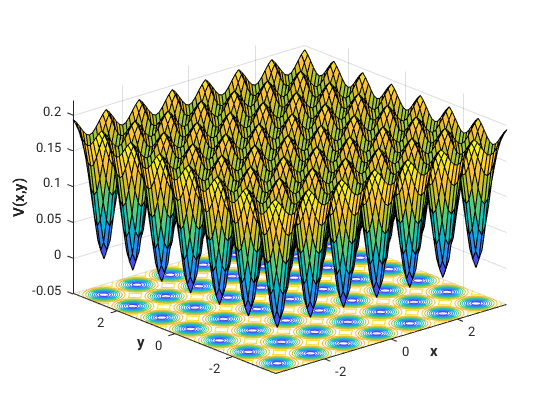}
\caption{The potential \eqref{metalpotentialanalytic} translated over a square ($N=7$) in each direction.  The potential minimum is set to zero making the corresponding Schr\"{o}dinger operator positive. 
}
\label{metalpotential}
\end{figure}

The final example for which we show the validity of this mechanism is a $2d$ solid periodic metal. In order to represent a realistic situation, we will not use potentials with vertical walls, but rather a smoothly varying function taken to be a Gaussian potential \cite{PavMar} assuming the form 
 \begin{align}\label{metalpotentialanalytic}
     V(x,y)=V_0e^{-(\alpha_x\frac{(x-x_0)^2}{a^2}+\alpha_y\frac{(y-y_0)^2}{a^2})},
 \end{align}
where $V_0$ represents the maximum depth of the well, $(x_0,y_0)$ are the coordinates of the center of the well, and $\alpha_x$ and $\alpha_y$ are measures of the range of the well in either direction. For the purpose of this paper we focus on the cases $\alpha_x=\alpha_y=a=1$ and $x_0=y_0=0$. The potential \eqref{metalpotentialanalytic} represents a single cell and is translated in each direction to model the overall potential of the metal (see Figure \ref{metalpotential}). The symmetry group corresponding to such potential is the discrete group $\mathbb{Z}_N$ acting in the obvious way (i.e. by addition) and $N^2$ denotes the total number of atoms.
In what follows we consider the $2d$ Schr\"{o}dinger operator induced by the potential \eqref{metalpotentialanalytic}, i.e.
\begin{align}\label{2dschroedingerbloch}
H_\hbar=-\hbar^2(\frac{d^2}{dx^2} + \frac{d^2}{dy^2})+V(x,y),
\end{align}
and we impose periodic boundary conditions and consider a finite domain. By Bloch's Theorem it follows that solutions to the eigenvalue problem $H_\hbar\psi_\hbar=E_\hbar\psi_\hbar$ exist and take the form of a plane wave modulated by a periodic function. In a similar fashion as Example \ref{schroedingerexample} and Example \ref{Curieweissexample} one can show that the algebraic vector state induced by the ground state eigenvector is $\mathbb{Z}_N$-invariant and localizes in each single well, as $\hbar\to 0$.

\begin{figure}[!htb]
\centering
\includegraphics[width=9cm,height=7cm]{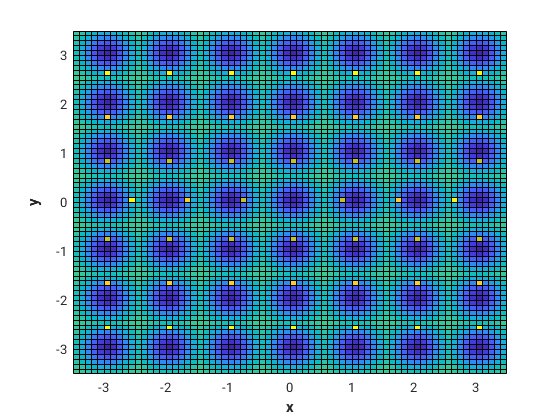}
\caption{Top view of the perturbed potential \eqref{metalpotentialanalytic} translated over $N=7$ cells in each direction. The dots in yellow indicate the support of the perturbation, each corresponding to a single grid point localized away from any of the minima.
}
\label{metalpotentialupper}
\end{figure}

To move on our discussion regarding localization we perturb the potential \eqref{metalpotentialanalytic} with a ``flea-like'' perturbation (cf. Property \ref{propertiesflealike}). In the same spirit as for the double well potential our perturbation is chosen to be strictly positive and bounded away from each of the minima. To avoid numerical complications we have taken the perturbation to be $\delta=0.1$ precisely on single grid points.  An upper view of the perturbed potential is depicted in Figure \ref{metalpotentialupper}.

\begin{figure}[!htb]
\centering
\includegraphics[width=15cm,height=7cm]{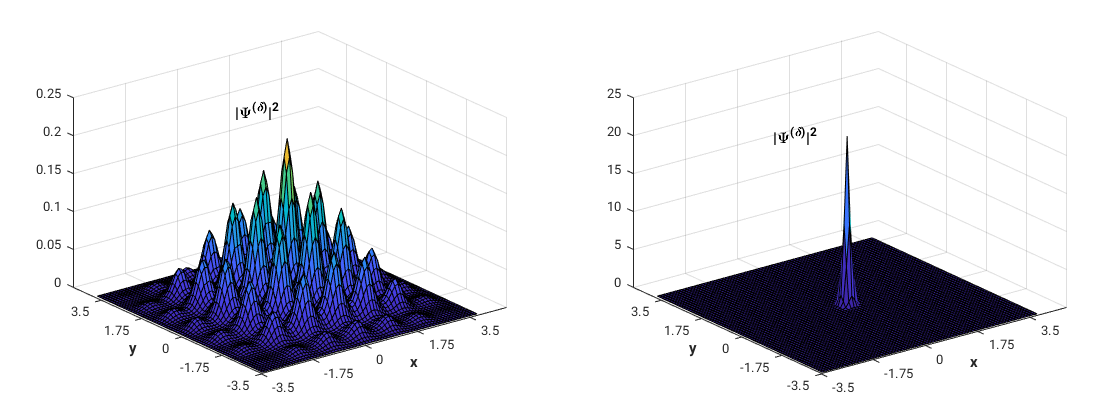}
\caption{The probability density for the ground state $\Psi_\hbar^{(\delta)}$ corresponding to the perturbed Schr\"{o}dinger operator for $\hbar=0.1$ (left) and $\hbar=0.025$ (right).  Little localization is observed for such relatively large values of $\hbar$ (left), whilst localization is almost total for such small values of $\hbar$ (right).
}
\label{groundstateperiodicflea}
\end{figure}

Similarly to the double well potential one can think this perturbation as having the property that the relative energy of the lattice is minimized exactly towards one cell, but the absolute minima are untouched. This is visible from Figure \ref{metalpotentialupper}, the center cell corresponding to $x=y=0$ is relatively lower in energy then all other cells since the perturbation is positive. It is clear from Figure \ref{groundstateperiodicflea} that the perturbed ground state localizes at the point $(0,0)$, as $\hbar\to 0$. Clearly the associated state (a Dirac measure) breaks the relevant symmetry.

\newpage
As before, it has been numerically checked that for the values of $\hbar$ used to compute the ground state eigenfunction $\Psi_\hbar^{(\delta)}$ of $H_\hbar^{(\delta)}$, the ground state energy $E_0^{(\delta)}$ is non-degenerate. Therefore, localization is not caused by ``numerical noise'' that typically does play a role when the (strictly speaking) non-degenerate eigenenergies differ by an order of magnitude comparable with the machine precision.

\begin{remark}
{\em 
We note that this perturbative approach generalizes Anderson's mechanism. Indeed, the ``flea''-mechanism that causes a pure state to arise in the classical limit has the same effect as the limit of the tower of states: it simply generates it!
}
\hfill$\blacksquare$
\end{remark}

\subsection{Paradox: case of phase transitions}\label{phase transitions paradox}
Despite the fact that Butterfield's Principle has been confirmed in the context of phase transitions, namely by viewing incipient singularities as robust behavior foreshadowing the singular behavior evidenced in infinite systems, the physical (and mathematical correct) idea of the ``flea'' illustrated above  has yet to be found.

Let us say a few words about the difficulties associated with phase transitions, characterized by coexisting states. Similar as for SSB, one may try to introduce suitable ``flea''-type perturbations that favor one of these states, in such a way that before the pertinent limit, this ``selection process'' already takes place. The main issue with phase transitions is that they might occur without the presence of additional symmetry and worse: all excited states may come into play. It is therefore a much harder issue to find suitable ``asymmetric'' perturbations.  To illustrate the latter, we limit ourselves to the $\mathbb{Z}_2-$invariant Curie-Weiss model (Example \ref{Curieweissexample}), but now for non-zero temperature $\beta$. In a similar fashion as elucidated in Section \ref{algclaslim} the relevant Gibbs state induced by the CW-Hamiltonian \eqref{CWham} converges to
\begin{align}
\omega_0^{(\beta)}(f)=\begin{cases}\frac{1}{2}(\delta_{(0,0,m_\beta/2}(f)+\delta_{(0,0,-m_\beta/2}(f)), \ \ (\beta> 4);\\
 \delta_{(0,0,0)}, \hspace*{4.25cm} (\beta\leq 4),
 \end{cases}
\end{align}
where $m_\beta$ is the temperature dependent magnetization, and $\delta_x(f)=f(x)$. The above probability measure is definitely $\mathbb{Z}_2-$invariant, in complete analogy to the case of SSB. Nonetheless, we expect Butterfield’s Principle to be restored through the introduction of asymmetric ``flea''-type perturbations to $H_N^{CW}$ that are localized in spin configuration space, although at nonzero temperature all excited states (rather than just the first) will start to play a role; the precise details of the ``flea'' scenario remain therefore to be settled.


\subsection{Comparison with the measurement problem}\label{Measurement problem}
In this last part, we outline the analogy between the flea mechanism and the measurement problem (we refer to \cite{LanReu13} for a detailed discussion). In broad terms, the measurement problem consists in the fact that the Schr\"{o}dinger equation of quantum mechanics fails to predict that measurements have single outcomes. Instead, it empirically predicts unacceptable ``superpositions'' thereof.
The obstacle here is that at first sight such super positions seem to survive the classical limit, as shown above by the ground state of the double-well potential in the limit $\hbar\to 0$ or by the ground state of the Curie-Weiss model in the limit $N\to\infty$. More generally, the measurement problem arises whenever in the classical limit a pure
quantum state converges to a mixed classical state, since in that case quantum theory fails to predict a single measurement outcome; it rather suggests there are many outcomes.  
Consequently, the real problem is to show that under realistic measurement conditions pure quantum states actually have pure classical limits.  As shown in \cite{LanReu13} this is brought about precisely by the ``flea''-mechanism. Indeed, In the presence of a small perturbation, the perturbed ground state will localize as $\hbar\to 0$, resulting in a pure state (Dirac measure).  Since this depends only on the support of the perturbation, but not on its size, the tiniest of perturbations may cause collapse in the classical limit. This ``collapse'' of the wave-function in the limit is thus compatible with the measurement problem.

\section{Discussion}
In this paper we have discussed the concept of spontaneous symmetry breaking and the theory of phase transitions from various points of view. 

Without going into technical details we have seen that the notion of the quantization map and more generally algebraic quantum theory is an  rigorous framework to study these natural phenomena. Even though mathematically correct, the results proved by such approaches do not always reflect reality; and in particular they violate Butterfield's Principle. Several methods and mechanisms we discussed are compatible with the physical idea of symmetry breaking in Nature, namely that the limiting state should be pure and not $G-$invariant. Probably the mechanism that best reflects this is the  ``flea-''mechanism. Indeed, a small perturbation of the Hamiltonian yields the  correct ``physical'' ground state  already {\em before} reaching the relevant regime which  therefore saves Butterfield's Principle. 
Without a mathematical proof, this mechanism has been generalized and confirmed by means of a simulation in case of a $2d$ periodic potential. 

However, it is not clear how such a flea should look like for more complex symmetries (e.g. continuous symmetry groups). Attempts have been made to introduce a similar flea perturbation in case of the Mexican hat potential but they did not seem to work: the perturbed ground state does not localize for any $\hbar$. In order to get localization, it is probably not sufficient to use a perturbation of the type \eqref{perturbationflea1}, i.e. its support is defined on larger sets. This raises the question whether or not such a perturbation is ``natural'' in the sense that it should resemble a tiny perturbation of the material. 

In the case of phase transitions at non-zero temperature the same issue remains: due to the in general absence of additional symmetry  and the interplay of {\em all} excited states it is not  not clear how to precisely define the flea ``scenario''.  It is therefore not clear whether this approach can resolve the paradox between Nature and Theory.

\section*{Acknowledgments}  
The research was financially supported by the Alexander von Humboldt Foundation. 
The author particularly thanks the referee for providing useful and fruitful comments, as well as Nicol\`{o} Drago for giving feedback on technical issues.

\end{document}